# Orientation Sensitive Terahertz Resonances Observed in Protein Crystals

Gheorghe Acbas[*], Edward Snell[#] and A.G. Markelz[*]

[*] Department of Physics, University at Buffalo, SUNY, Buffalo, New York, USA
[#] Department of Structural Biology, University at Buffalo, SUNY, Buffalo, New York, USA
Email: amarkelz@buffalo.edu

*Abstract—* A method is presented for measuring anisotropic THz response for small crystals, Crystal Anisotropy Terahertz Microscopy (CATM). Sucrose CATM measurements find the expected anisotropic phonon resonances. CATM measurements of protein crystals find the expected broadband water absorption is suppressed and strong orientation and hydration dependent resonant features.

## I. Introduction And Background

CORRELATED motions in proteins have long been predicted to lay in the terahertz frequency range [1]. Unfortunately the measurement of these modes has been problematic due to overlap with the broadband response of biological water and possible librational motions of surface side chains. Polarization difference spectroscopy is a method that can be used to suppress a homogeneous background from orientation sensitive resonances, however the size for typical protein crystals is far below the diffraction limit at terahertz frequencies. Recently great advances in THz near field microscopy have been made[2], however typical systems do not focus on the spectroscopic quality of the measurement, or the monitoring of anisotropic response. Here we demonstrate a method for measuring small crystals (~300 μm) in the terahertz range, and find strong orientation sensitive features from protein crystals. We refer to our technique as Crystal Anisotropy Terahertz Microscopy (CATM).

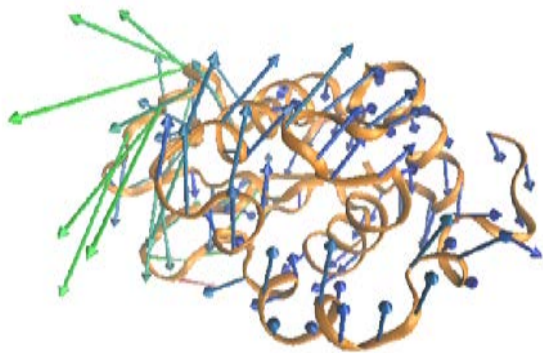

**Figure 1. Displacement vector diagram for lowest vibrational mode of HEWL, 4.5 cm$^{-1}$ = 0.13 THz.**

Dynamic motions of complexes are universal in biology, encompassing processes as diverse as mitosis, signal transduction, regulatory pathways, immune response, protein folding, and enzymatic activity [3]. For molecular recognition, for example, proteins adapt their structure to different binding partners, often exhibiting large structural changes. The transition in these molecules can be modeled, determining a low energy pathway of collective motion. Calculations show that functional conformational change in many biomolecular systems can be simulated using only the first few collective vibrational modes of the system [4, 5]. While it is possible that large scale conformational change occurs through diffusive configurational sampling, concerted fluctuations could explain the observed high physiological on-rates and affinities [6-9].

These correlated motions are the low frequency vibrational modes that extend throughout the macromolecule. The motions can be estimated through normal mode calculations. Here we use CHARMM23 and the protein structure file for hen egg white lysozyme (HEWL) 1bwh.pdb to calculate the vibrational density of states and the estimated absorption. In Fig. 1 we show the displacement vector diagram for the lowest frequency mode. The calculation includes the crystal waters and the lowest vibrational mode frequency was calculated as 4.5 cm$^{-1}$ or 0.13 THz. The displacement vectors are drawn from the α carbons and their lengths are proportional to the atomic displacement for the net eigenvector [10]. It has long been speculated that these correlated motions, while at the picosecond time scale, contribute to protein conformational dynamics and alteration of the motions can effect function. There are currently no direct measurements of these motions within the standard biophysical toolbox. NMR can give information about local motions, but cannot indicate correlation of motions. Similary X-ray crystallography measures regions of flexibility through the Debye Waller factor, but cannot identify specific modes contributing to the net atomic mean squared displacement.

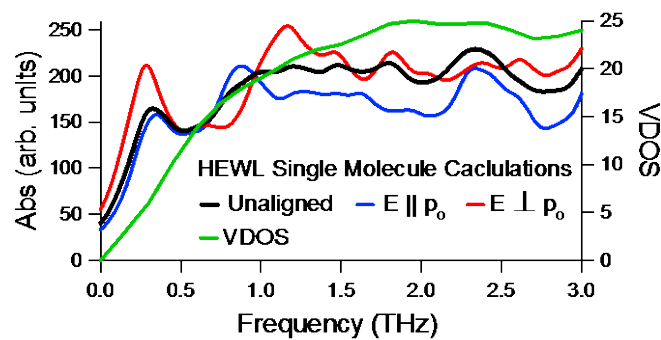

**Figure 2. Calculated VDOS and absorbance for unaligned and aligned HEWL.**

Spectroscopy measurements, in particular inelastic scattering methods such as X-ray inelastic scattering and neutron inelastic, have shown promise for measuring and identifying modes [11, 12]. These techniques are particularly appealing as they can map out the full phonon dispersion. However these are facility based measurements and require large sample sizes and deuteration. Terahertz spectroscopic measurements have been attempted, however the challenge is to observed specific modes. In Fig. 2 we show the calculated vibrational density of states (VDOS), showing that a high density in the THz range. In the same figure we show the calculated absorbance for an unaligned sample, corresponding to bulk powder or solution phase samples.

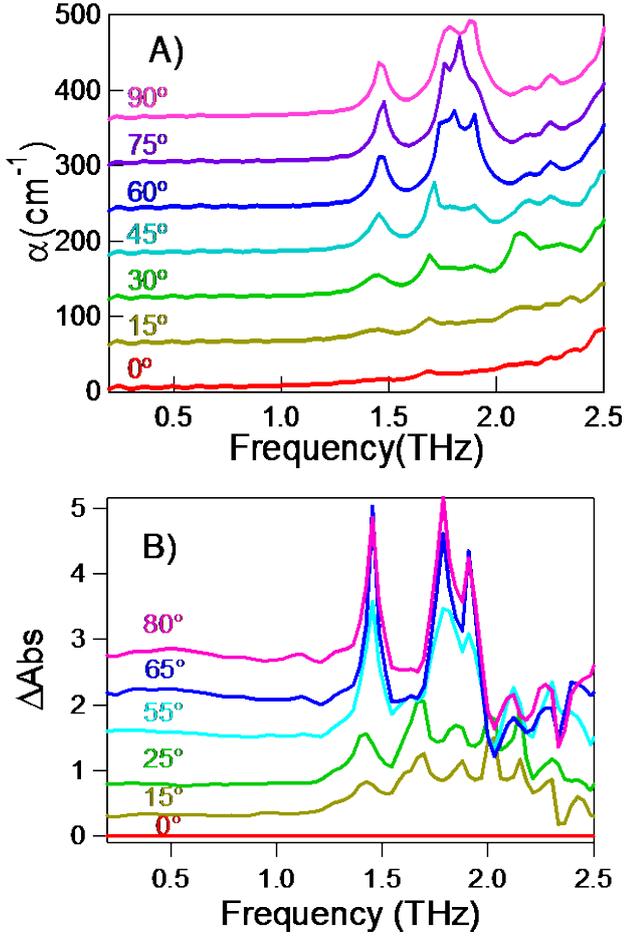

Figure 3. Demonstration of CATM for a sucrose crystal. A) Far field measurement of an a-face polished sucrose crystal measured with standard far field THz TDS. B) single pixel near field measurement of sucrose crystal measured with CATM. 90° corresponds to E||b.

Even when one includes the dipole coupling of the modes, the spectrum is still very smooth. In addition, librational motion of bound water and the polypeptide sidechains will give rise to a broad relaxational background, that may dominate the response, thus a typical terahertz spectroscopic measurement of protein samples give a broad featureless absorbance without any apparent dominant modes. However one might achieve high mode constrast if one considers A) the relaxational contribution should be isotropic with respect to the protein structure and B) dipole coupling to protein structural modes will be anisotropic. Also shown in Fig. 2 are the calculated absorbance for aligned samples, where the incident THz polarization is parallel and perpendicular to the static dipole of the protein. As seen in the figure, the spectrum is not as smooth as the unaligned sample. However if one simply measures the THz transmission for an aligned sample, the relaxational contribution will still dominate the measured absorbance. By taking polarization difference measurements one should be able to eliminate the relaxational contribution and enhance contrast between modes. A naturally aligned system is a crystal, however protein crystals are rarely larger than a few hundred micron in size, below the diffraction limit of THz measurements. A careful set of measurements has been reported using THz time domain spectroscopy (THz TDS) of HEWL crystals to investigate if narrow band absorption could be observed [13]. Measurements were made in the far field as a function of hydration. Unfortunately the only narrow band features observed in these measurements were due to changing atmospheric water content. Orientation dependence was not attempted in those measurements. It is possible that diffraction effects smoothed the strong absorbances that we report here.

To overcome diffraction limitations we use a THz TDS near field microscope method based on that by Planken [2]. This approach uses standard photoconductive THz generation and electro-optic detection, however the sample is placed directly on top of a horizontal electro optic crystal and the resolution is set by the spot size of the NIR probe incident from the back of the crystal. In our case the ZnTe crystal forms part of a humidity controlled chamber for the sample. The hydration in the chamber is controlled by flow from a Licor Dewpoint generator. A THz image is formed by scanning the ZnTe-sample stage with a THz waveform measured for each pixel. The spatial resolution of our system is 10-30 μm. The THz system is enclosed and purged with dry nitrogen. All measurements were performed at room temperature.

Samples were mounted on top of a thin aluminum plate (~200-300 μm thick) in which small apertures have been drilled (200-300 μm diameter) [14]. Different orientations are measured by rotating the plate. A THz image is taken of the aperture for each orientation. The spectra from the center pixels for each orientation are compared. A challenge in measuring the absorbance is proper referencing and removal of etalon. The metal aperture has waveguide transmission characteristics, with a cutoff that is dependent on dielectric filling, complicating the use of an empty aperture as a reference. Molecular crystal measurements using single-mode parallel plate waveguide with the crystalline films grown on the waveguide walls did not require any reference as the strong narrow resonant features were readily apparent in the power spectrum and there is no concern for etalon, as the sample region is long with no apparent reflected pulse artifacts.[15]. In the CATM measurements however, transmission is through a ~ 300 um thick crystal, thus we expect reflection at the interfaces that cannot be removed from the main pulse. We address both the referencing and etalon concerns by self-referencing. As we are interested in the change in absorbance with orientation, we use a single orientation of the crystal as our reference and calculate a difference absorbance using the following:

$$\Delta Abs = -2\ln\left[\frac{|E_t(\omega,\theta)|}{|E_t(\omega,\theta_{ref})|}\right]$$
$$= -2\ln\frac{F(\omega)|E_i(\omega)|e^{-\alpha(\omega,\theta)d/2}}{F(\omega)|E_i(\omega)|e^{-\alpha(\omega,\theta_{ref})d/2}} \quad (1)$$
$$= \left[\alpha(\omega,\theta) - \alpha(\omega,\theta_{ref})\right]d$$

where $|E_t(\omega,\theta)|$ ($|E_i(\omega,\theta)|$) is the magnitude of the transmitted (incident) electric field, $\alpha(\omega,\theta)$ is the sample's absorption at orientation angle $\theta$, and $d$ is the sample thickness. $F(\omega)$ is the frequency dependent transmission due to Fresnel loss at interfaces, waveguide transmission for the aperture, and etalon effects. This factor should be orientation independent. By using the center pixels of the aperture as a given orientation spectral measurement, we ensure that the thickness of the sample is always the same for the faceted crystal, thus the removal of etalon should be valid as long as there is no strong birefringence. We show in Fig. 3 that self-referencing works well for sucrose which has a birefringence of $\Delta n = 0.05$.

Calibration and characterization measurements were made with a sucrose crystal, which has well known birefringence and anisotropic absorbance. Large sucrose crystals were grown and polished to ~500 μm thickness. These crystals were measured both using far field standard THz TDS as well as near field CATM.

Lysozyme crystals were prepared using the sitting drop method. The protein was supplied by Sigma-Aldrich and used at 60 mg/ml in 0.1M NaAc pH 5.2 buffer without further purification. The precipitant used was 10% NaCl in the same buffer. The sitting drop contained 10 μl of precipitant and 10 μl of protein solution with 500 μl of precipitant in the reservoir. Crystals grew over a period of a few days. Tetragonal hen egg white lysozyme crystals are mounted on the sample plate and covered with paraffin oil to maintain hydration and then placed in the ZnTe chamber.

## II. RESULTS

Sucrose crystal measurements demonstrate CATM. Fig. 3 shows the anisotropic absorbances measured with CATM are in agreement with the far field measurements for large crystals.

In Fig. 4 we show the orientation dependence of the difference spectra relative to 0° rotation for a hydrated lysozyme crystal. As seen in the Figure there are three clear orientation dependent resonances at 40 cm-1, 60 cm-1 and

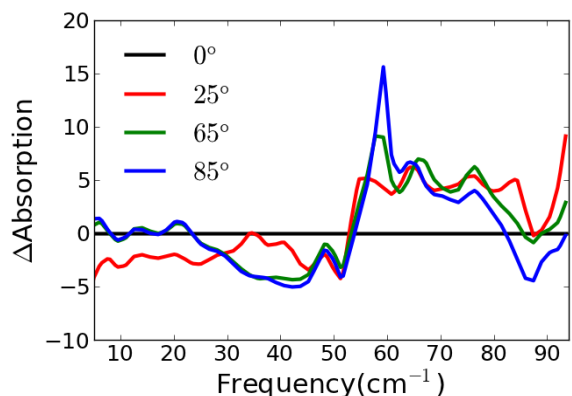

**Figure 4.** Absorption difference spectra of hydrated HEWL crystal for several orientations.

85 cm$^{-1}$. These strong features cannot arise from typical artifacts, as these effects should be removed by self-referencing. We note that since we are referencing to a single orientation, negative resonances indicate that the resonance is present in the reference orientation and is decreasing in the other orientations. Water and relaxational absorbances are also removed by the self-referencing, however even in the net signal we do not see the expected bulk water absorbance. This lack of relaxational background is suprising given that the the hydration for the crystals is greater than many hydrated powder and film measurements which measure a large background.

## III. CONCLUSION

Orientation dependent sharp absorbances are observed for sucrose and protein crystals using the CATM technique. The orientation dependent features seen in sucrose are in agreement with far field measurements, demonstrating the technique gives reliable anistropic spectral data. The absorbances are likely related to correlated motions within the protein molecules.

ACKNOWLEDGMENT

We thank the National Science Foundation MRI^2 grant DBI2959989 for support.